\documentclass[12pt]{article}
\begin{document}

\bigskip

\begin{center}
{\Large \bf  ON TWO NONINTEGRABLE CASES OF THE
         GENERALIZED H\'ENON--HEILES SYSTEM 
	 WITH AN ADDITIONAL NONPOLYNOMIAL TERM\\[7.2mm]
E.~I.~Timoshkova* and S.~Yu.~Vernov**} \\[2.7mm]
*{\it Central Astronomical Observatory at Pulkovo\\
Pulkovskoe sh. 65/6, Saint-Petersburg, 196140,  Russia}\\
E-mail: elenatim@gao.spb.ru\\[2.7mm]
{*}*{\it Skobeltsyn Institute of Nuclear Physics, Moscow
State University,\\
Vorob'evy Gory, Moscow, 119992,  Russia}\\
E-mail: svernov@theory.sinp.msu.ru
\end{center}

\bigskip

\begin{abstract}
\normalsize The generalized H\'enon--Heiles system with an additional nonpolynomial term
is considered.  In two nonintegrable cases new two-parameter solutions
have been obtained in terms of elliptic functions.  These solutions
generalize the known one-parameter solutions.  The singularity analysis
shows that it is possible that three-parameter single-valued solutions exist
in these two nonintegrable cases. The knowledge of the Laurent series
solutions simplifies search of the elliptic solutions
and allows to automatize it.
\end{abstract}

\bigskip


\section{INTRODUCTION}

 Beginning from papers~[1--3], investigations
of two-dimensional Hamiltonian systems with polynomial potentials
attract large attention due to detect of the "dynamical chaos"
phenomena. There is no method to find the multivalued 
general solution of a two-dimensional nonintegrable system in the analytic 
form. At the same time it is an actual problem to find single-valued 
special solutions in the analytic form, because the investigation of the 
solutions with some additional properties, for example, periodic 
solutions, plays an important role in the study of physical phenomena.  
Another problem is to pick out nonintegrable cases, in which single-valued 
special solutions can depend on maximal number of arbitrary parameters.

 The H\'enon--Heiles Hamiltonian~\cite{HeHe}:
\[
H=\frac{1}{2^{\vphantom{27}}}\bigl(x_t^2+y_t^2+x^2+y^2)+x^2y-
\frac{1}{3^{\vphantom{27}}}
\:y^3
\]
and its generalizations are one of the most actively studied
two-dimensional Hamiltonians (see~\cite{Vernov0302} and references
therein). The generalized H\'enon--Heiles system is a model widely used
in astronomy~\cite{Murray}  and physics, for
example, in gravitation~\cite{Kokubun,Podolsky}.

One of lines of investigation of this system is the search for special
solutions~[8--13].
The general solutions in the analytic form are known only in the
integrable cases~[14--17],
in other cases not
only four-, but even three-parameter exact solutions have yet to be found.
In~\cite{Timosh} new type of one-parameter elliptic solutions has been
obtained. Such solutions exist only in integrable cases and two
nonintegrable ones. In these nonintegrable cases there exist
three-parameter Laurent-series solutions~\cite{VernovTMF},
which generalize the Laurent series of one-parameter elliptic solutions.
 In this paper we find elliptic
two-parameter solutions, which generalize solutions
obtained in~\cite{Timosh}.

\section{BASIC EQUATIONS}

The generalized H\'enon--Heiles system with an additional nonpolynomial
term is described by the Hamiltonian
$$
H=\frac{1}{2}\Big(x_t^2+y_t^2+\lambda_1
x^2+\lambda_2 y^2\Big)+x^2y-\frac{C}{3}\:y^3+\frac{\mu}{2x^2} \eqno(1)
$$
and the corresponding system of the motion equations:
$$
  \left\{ \begin{array}{lcl}
  \displaystyle x_{tt}^{\vphantom{7}} {}={}-\lambda_1 x
-2xy+\frac{\mu}{x^3},\\[2mm]
 \displaystyle  y_{tt}^{\vphantom{7}} {}={}-\lambda_2 y -x^2+Cy^2,
\end{array}
\right.
\eqno(2)
$$
where $x_{tt}^{\vphantom{7}}\equiv\frac{d^2x}{dt^2}$ and
$y_{tt}^{\vphantom{7}}\equiv\frac{d^2y}{dt^2}$,  $\lambda_1$, $\lambda_2$,
$\mu$ and $C$
are arbitrary numerical parameters. Note, that if  $\lambda_2\neq 0$,
then one can put  $\lambda_2=sign(\lambda_2)$ without the loss of
generality.

Due to the Painlev\'e analysis~[19--21] the 
following integrable cases have been found~\cite{Tabor}:  \[ \begin{array} 
{cll} \mbox{(i)} & C=-1, &\lambda_1=\lambda_2,\\ \mbox{(ii)} & C=-6, 
&\mbox{$\lambda_1$, $\lambda_2$ arbitrary,}\\ \mbox{(iii)} & C=-16,\quad
&\lambda_1=\lambda_2/16.\\
\end{array}
\]

These integrable cases correspond precisely to the stationary flows of the
only three integrable cases of the fifth-order polynomial nonlinear
evolution equations of scale weight 7 (respectively the Sawada--Kotega,
the fifth-order Korteweg--de Vries and the Kaup--Kupershmidt
equations)~\cite{Weiss1,Polish}.

In all above-mentioned cases system $(2)$ is integrable at any
value of $\mu$. The function $y$, solution of system~$(2)$, satisfies
the following fourth-order equation~\cite{CoMu92,Timosh,Polish}:
$$
y_{tttt}^{\vphantom{7}}=(2C-8)y_{tt}^{\vphantom{7}}y
- (4\lambda_1+\lambda_2)y_{tt}^{\vphantom{7}}+2(C+1)y_{t}^2+
\frac{20C}{3}y^3+ (4C\lambda_1-6\lambda_2)y^2-4\lambda_1\lambda_2 y-4H,
\eqno(3)
$$
where $H$ is the energy of the system. We note, that $H$ is
not an arbitrary parameter, but a function of initial
data: $y_0^{\vphantom{7}}$, $y_{0t}^{\vphantom{7}}$,
 $y_{0tt}^{\vphantom{7}}$ and $y_{0ttt}^{\vphantom{7}}$. The
form of this function depends on $\mu$:
$$
H=\frac{1}{2}(y_{0t}^2+y_0^2)-\frac{C}{3}y_0^3+
\left(\frac{\lambda_1}{2}+y_0^{\vphantom{7}}\right)(Cy_0^2-
\lambda_2y_0^{\vphantom{7}}-y_{0tt}^{\vphantom{7}})+
\frac{(\lambda_2y_{0t}^{\vphantom{7}}+
2Cy_0^{\vphantom{7}}y_{0t}^{\vphantom{7}}-y_{0ttt}^{\vphantom{7}})^2+\mu}
{2(Cy_0^2-\lambda_2y_0^{\vphantom{7}}-y_{0tt}^{\vphantom{7}})}.
$$

This formula is correct only if $x_0^{\vphantom{7}}=Cy_0^2
-\lambda_2y_0^{\vphantom{7}}-y_{0tt}^{\vphantom{7}}\neq0$.
If $x_0^{\vphantom{7}}=0$, what is possible only at $\mu=0$, then we can 
not express $x_{0t}^{\vphantom{7}}$ through $y_0^{\vphantom{7}}$, 
 $y_{0t}^{\vphantom{7}}$, $y_{0tt}^{\vphantom{7}}$ and 
$y_{0ttt}^{\vphantom{7}}$, so  $H$ is not a function of the initial data.  
If $y_{0ttt}^{\vphantom{7}}
=2Cy_0^{\vphantom{7}}y_{0t}^{\vphantom{7}}-\lambda_2y_{0t}^{\vphantom{7}}$, 
then eq. $(3)$ with an arbitrary $H$ corresponds to system $(2)$ with
$\mu=0$, in opposite case
eq. $(3)$ does not correspond to system $(2)$.

To find a special solution of
eq.~$(3)$ one can assume that $y$ satisfies some more simple
equation. For example, there exist solutions
in terms of the Weierstrass elliptic functions, which satisfy the
following equation:
$$
y_t^2={\cal A}y^3+{\cal B}y^2+{\cal C}y+{\cal D}, \eqno(4)
$$
where
${\cal A}$, ${\cal B}$,
${\cal C}$ and ${\cal D}$ are some constants.

The following generalization of eq.~$(4)$:
$$
  y_t^2=\tilde {\cal  A}y^3+\tilde {\cal B} y^{5/2}+\tilde {\cal C} y^2+
\tilde  {\cal D} y^{3/2}+\tilde {\cal E} y+\tilde {\cal G}
\eqno(5)
$$
gives new one-parameter solutions in two nonintegrable cases~\cite{Timosh}:
  $C=-16/5$ and  $C=-4/3$  ($\lambda_1$ is an arbitrary number,
$\lambda_2=1$).
It is easy to show~\cite{Timosh} that if $\tilde {\cal B}\neq 0$ or
$\tilde {\cal D} \neq 0$ then $\tilde {\cal G}=0$,
therefore, substitution $y=\varrho^2$ transforms eq.~$(4)$ into
$$
\varrho_t^2=\frac{1}{4}\Bigl(\tilde {\cal A}\varrho^4+\tilde {\cal B}
\varrho^3+\tilde  {\cal C} \varrho^2 + \tilde {\cal D}\varrho+\tilde {\cal
E}\Bigr).  \eqno (6)
$$
In~\cite{Timosh2} using the substitution $y\longrightarrow y-P_0$
a new parameter $P_0$ has been introduced and two-parameter solutions
have been constructed for above-mentioned values of $C$ and a few values
of $\lambda_1$ ($\lambda_2=1$). Due to Painlev\'e analysis local
three-parameter solutions as the converging Laurent series have been found
for an arbitrary $\lambda_1$, $\lambda_2=1$ and $\mu=0$~\cite{VernovTMF}.
In the present paper we seek both the elliptic and the Laurent-series 
solutions for arbitrary values of $\lambda_1$, $\lambda_2$ and $\mu$.

\section{NEW SOLUTIONS}
    Let us assume that solutions of eq.~$(3)$ in the neighborhood of
singularity point $t_0$ tend to infinity as $y=c_{\beta}(t-t_0)^\beta$,
where $\beta$ and  $c_{\beta}$ are some complex numbers.
Of course, the real part of $\beta$ has to be less then zero.
From this assumption it follows~\cite{Tabor} that $\beta=-2$.
The Laurent series of solutions of eq.~$(6)$ begin with  term
proportional to $(t-t_0)^{-1}$, so we seek solutions of eq.~$(3)$ as
square polynomial: $y=P_2\varrho^2+P_1\varrho+P_0$, where $P_2$, $P_1$ and
$P_0$ are arbitrary numbers,  $\varrho$ is the general solution of
eq.~$(6)$ with arbitrary coefficients  $\tilde {\cal A}$, $\tilde {\cal
B}$, $\tilde  {\cal C}$, $\tilde {\cal D}$ and $\tilde {\cal E}$.
Because of the function $\tilde\varrho=(\varrho-\frac{P_1}{2})/\sqrt{P_2}$
is a solution of eq.~$(6)$ as well, we can put $P_2=1$ and $P_1=0$
without loss the generality. 

Substituting $y=\varrho^2+P_0$ in eq.~$(3)$, we obtain
$$
\begin{array}{l}
\displaystyle \varrho_{tttt}\varrho {}= {} - 4\varrho_{ttt}\varrho_t
 - 3\varrho_{tt}^2
+ 2(C-4)\varrho_{tt}\varrho^3 + (2P_0(C-4)-4\lambda_1
-\lambda_2)\varrho_{tt}\varrho
+ { } \\ \displaystyle  { } + 2(3C-2)\varrho_{t}^2\varrho^2 +
(2CP_0-4\lambda_1-8P_0-\lambda_2)\varrho_{t}^2+
\frac{10}{3}C\varrho^6 + { }
\\ \displaystyle  { } + (2C\lambda_1 + 10CP_0- 3\lambda_2)\varrho^4 +
2(2\lambda_1CP_0 +5CP_0^2 -\lambda_1\lambda_2- 3P_0\lambda_2)\varrho^2  +
{ } \\ \displaystyle  { } + \frac{10}{3}CP_0^3 + 2\lambda_1 CP_0^2  -
3P_0^2\lambda_2 - 2\lambda_1\lambda_2 P_0- 2H.  \end{array}
\eqno(7)
$$

The function $\varrho$ is a solution of eq.~$(6)$, hence,
eq.~$(7)$ is equivalent to the following system:
$$
\left\{
\begin{array}{l}
     (3{\tilde\mathcal{A}} + 4)\,( - 3{\tilde\mathcal{A}} + 2C)=0,\\
{\tilde\mathcal{B}} ( - 21 {\tilde\mathcal{A}} + 9 C - 16)=0, \\
 96 {\tilde\mathcal{A}} C P_0  - 240 {\tilde\mathcal{A}} {\tilde\mathcal{C}} -
 192 {\tilde\mathcal{A}} \lambda_1 - 384 {\tilde\mathcal{A}}
P_0 - 48 {\tilde\mathcal{A}}\lambda_2  - { } \\ { }
- 105 {\tilde\mathcal{B}}^2 +
128 {\tilde\mathcal{C}} C - 192 {\tilde\mathcal{C}} + 128 C \lambda_1 +
640 C P_0 - 192\lambda_2 =0,\\
 40 {\tilde \mathcal{B}} C P_0   - 90 {\tilde\mathcal{A}}
  {\tilde\mathcal{D}} - 65 {\tilde\mathcal{B}} {\tilde\mathcal{C}} - 80
{\tilde\mathcal{B}}  \lambda_1 - 160 {\tilde\mathcal{B}} P_0 - 20
{\tilde\mathcal{B}}\lambda_2 + 56 C{\tilde\mathcal{D}}  - 64
{\tilde\mathcal{D}}=0,\\ 16 {\tilde\mathcal{C}} C P_0 - 36
{\tilde\mathcal{A}} {\tilde\mathcal{E}} - 21 {\tilde\mathcal{B}}
{\tilde\mathcal{D}}  - 8 {\tilde\mathcal{C}}^2 - 32 {\tilde\mathcal{C}}
 \lambda_1 - 64 {\tilde\mathcal{C}} P_0 - 8\lambda_2 {\tilde\mathcal{C}} +
24 C {\tilde\mathcal{E}}+ { } \\ { }  + 64 \lambda_1CP_0 + 160 C P_0^2 -
 16 {\tilde\mathcal{E}} - 32 \lambda_1\lambda_2 - 96 P_0\lambda_2=0,\\ 10
{\tilde\mathcal{B}} {\tilde\mathcal{E}} + (5 {\tilde\mathcal{C}} + 8CP_0 -
16\lambda_1 -  32P_0 - 4\lambda_2){\tilde\mathcal{D}}=0 ,\\ 384 H = - 48
  {\tilde\mathcal{C}} {\tilde\mathcal{E}} + 96 C {\tilde\mathcal{E}} P_0 +
 384 C \lambda_1 P_0^2 + 640 C P_0^3 - 9 {\tilde\mathcal{D}}^2- { }\\ { }
- 192 {\tilde\mathcal{E}} \lambda_1 - 384 {\tilde\mathcal{E}} P_0 - 48
{\tilde\mathcal{E}}\lambda_2 - 384 \lambda_1\lambda_2 P_0 -
576\lambda_2P_0^2.\\ \end{array} \right.
\eqno(8)
$$
System~$(8)$ has been solved by computer algebra software
REDUCE~\cite{REDUCE}.

If ${\tilde{\cal B}}\neq 0$, then from two first equations of
system~$(8)$ we obtain:
$$ C=-\:\frac{4}{3}  \quad\mbox{and}\quad
{\tilde\mathcal{A}}=-\:\frac{4}{3}\qquad \mbox{or} \qquad
C=-\:\frac{16}{5} \quad\mbox{and}\quad
{\tilde\mathcal{A}}=-\:\frac{32}{15}.
$$

 If ${\tilde{\cal B}}=0$, then solutions with ${\tilde{\cal
 D}}\neq 0$ are also possible at $C=-16$ and $C=-1$, but only in
 integrable cases.
The obtained solutions of eq.~$(3)$ depend on two parameters:
energy $H$ expressed through $P_0$ and parameter $t_0$
connected to homogeneity of time.

Six solutions of system $(8)$ correspond to each value of $P_0$.
Two of them (with $\tilde\mathcal{B}=\tilde\mathcal{D}=0$)
generate solutions of eq. $(4)$.
Values of ${\tilde\mathcal{B}}$ and ${\tilde\mathcal{D}}$, corresponding
to other solutions, depend on $\lambda_1$ and $\lambda_2$ and 
are zero only at some relations between these parameters.
We will consider only solutions with
${\tilde\mathcal{B}}\neq 0$ or ${\tilde\mathcal{D}}\neq
0$. They are presented in Appendix. These solutions can be
separated on pairs in such a way that solutions in one pair  differ
only in signs of $\tilde\mathcal{B}$ and $\tilde\mathcal{D}$.  Basic
properties of the obtained solution are considered in this section. In the
next section we analyze in detail solutions of system $(8)$ for some
values of $\lambda_1$ and $\lambda_2$.

   If the right-hand side of eq.~$(6)$ is a polynomial with multiple
  roots, then $\varrho$ and $y$ can be expressed in terms of elementary
  functions.  In opposite case $y$ is an elliptic function~\cite{BE,Hurwitz}.

It is simplicity itself that 
$y(t)=\varrho^2(t)+P_0=(-\varrho(t))^2+P_0$, 
 so, solutions of system $(8)$ with
opposite values of  $\tilde{\cal B}$ and $\tilde{\cal D}$ generate
identical solutions of eq.~$(3)$. 
From eq.~$(6)$ we obtain a polynomial equation for $y(t)$:
$$
(y_t^2-\tilde\mathcal{A}(y-P_0)^3-
\tilde\mathcal{C}(y-P_0)^2-\tilde\mathcal{E}(y-P_0))^2
=(y-P_0)^3(\tilde\mathcal{B}(y-P_0)+\tilde\mathcal{D})^2.   \eqno(9)
$$

The function $\varrho(t)$
can be expressed through the Weierstrass elliptic
function $\wp(t)$~\cite[Ch.~5]{Hurwitz}:
$$
\varrho(t-t_0)=\frac{a\wp(t-t_0)+b}{c\wp(t-t_0)+d}, \qquad (ad-bc=1),
$$
where $t_0$ is an arbitrary parameter. Periods of
$\wp(t)$ and the constants $a$, $b$, $c$ and $d$ are determined by
eq.~$(6)$.  The function
$$
y(t-t_0)=\left(\frac{a\wp(t-t_0)+b}{c\wp(t-t_0)+d}\right)^2+P_0 \eqno(10)
$$
is the fourth-order elliptic function. This function, as a solution of
eq.~$(3)$, can have only the second-order poles, therefore, in the
parallelogram of periods it has two poles with
opposite residues.  Solutions $(10)$ differ from solutions of eq.~$(4)$,
which are the second-order elliptic functions~\cite{Hurwitz}.

The function $x(t)$ satisfies the first equation of system~$(2)$ with
$$
\begin{array}{@{}l@{}}
\displaystyle \mu=
 \frac{8}{3}C^2{P_0}^5+\left(2\lambda_1C^2
-\frac{14}{3}\lambda_2C\right)P_0^4+ \left(2\lambda_2^2
-\frac{10}{3}C\tilde\mathcal{E}-4\lambda_1\lambda_2C\right)P_0^3
+ \left( 2\lambda_1\lambda_2^2 - { }\right. \\[2.7mm]
\displaystyle{} - \left.2\lambda_1C\tilde\mathcal{E}
-4CH+ 3\lambda_2\tilde\mathcal{E}\right)P_0^2 +\left(
2\lambda_1\lambda_2\tilde\mathcal{E}+\tilde\mathcal{E}^2 +4\lambda_2H 
\right) P_0+2\tilde\mathcal{E}H+\frac{1}{2} 
\lambda_1\tilde\mathcal{E}^2+\frac{9}{128} \tilde\mathcal{D}^2\tilde\mathcal{E}.
\end{array}\eqno(11)
$$

The trajectory of the
motion can be derived from the second equation of system $(2)$.
Substituting $y_{tt}^{\vphantom{7}}$, we obtain:
$$
x^2=(C-\frac{3}{2}\tilde {\cal  A})y^2+
  (3\tilde{\cal A}P_0-\tilde{\cal C}-1)y - \frac{1}{4}(5\tilde {\cal B}
y+3 \tilde {\cal D}- 5\tilde{\cal B}P_0)\sqrt{y-P_0}-
\frac{1}{2}(\tilde {\cal E} +3\tilde {\cal A} P_0^2- 2\tilde {\cal C}
  P_0).  
$$

If $\tilde {\cal B}$ and $\tilde {\cal D}$ take
zero values we get simple algebraic
trajectories. The full list of such trajectories
is presented in~\cite{AnTi}.  The parameter $P_0$ is absent in
these trajectory equations.

One value  of the energy $H$ can correspond to no more than three values
of $P_0$ and, hence, no more than six different one-parameter solutions.
Solutions $(10)$ differ from solutions of eq.~$(4)$, which are
the second-order elliptic functions~\cite{Hurwitz}.

\section{A PARTICULAR CASE}

\subsection{The form of solutions}

At $C=-16/5$,  $\lambda_1=1/9$ and $\lambda_2=1$ one-parameter solutions
($P_0=0$)   have been considered in detail in our previous
papers~\cite{Timosh, VernovTMF}. For these values of parameters
solutions of system $(8)$ are:
\[ \begin{array}{@{}llll@{}} {\bf 1.} &
\displaystyle\tilde\mathcal{A}={\displaystyle -\:\frac{32}{15}}, &
\displaystyle\tilde\mathcal{B}=0, & \displaystyle\tilde\mathcal{C}=
-\:\frac{32}{5} P_0 - 1, \\[2.7mm] &\displaystyle\tilde\mathcal{D}=0, &
\displaystyle\tilde\mathcal{E}= - \:\frac{32}{5}P_0^2 - 2P_0, \quad &
\displaystyle H=\frac {16}{15}  P_0^{3} +
 \frac {1}{2}  P_0^{2}\\[4.7mm]
{\bf 2.} &\displaystyle\tilde\mathcal{A}=-\:\frac{4}{3},\quad  &
\displaystyle\tilde\mathcal{B}=0,\quad  &
\displaystyle\tilde\mathcal{C}= -4P_0 - \frac {17}{33},\\[2.7mm]
&\displaystyle\tilde\mathcal{D}=0, &
\displaystyle\tilde\mathcal{E}= -4P_0^{2} - \frac {34}{33} P_0 + \frac
{20}{3267}, \quad& \displaystyle H= -\:\frac {2}{15}  P_0^{3} - \frac
{17}{330}P_0^{2} + \frac{2}{3267}P_0 - \frac{230}{323433},\\[3.7mm]
\end{array}
\]
\[
\begin{array}{@{}llll}
{\bf 3-4.} &\displaystyle\tilde\mathcal{A}=-\:\frac{32}{15},\quad &
\displaystyle\tilde\mathcal{B}=\pm \frac{8i\sqrt{15}}{45},\quad  &
\displaystyle\tilde\mathcal{C}= -\:\frac{32}{5}P_0 - \frac{4}{9},\\[2.7mm]
&\displaystyle \tilde\mathcal{D}=\pm\frac{4i\sqrt{15}}{9}P_0,\qquad   &
\displaystyle \tilde\mathcal{E}= - \frac{32}{5}
P_0^{2} - \frac{8}{9}P_0, \qquad &\displaystyle  H= \frac{16}{15}P_0^{3} -
\frac{7}{72}P_0^{2}, \qquad \qquad \qquad \quad { } \\[3.7mm]
\end{array}
\]
\[
\begin{array}{@{}lll@{}}
{\bf 5-6.} &\displaystyle\tilde\mathcal{A}= -\:\frac{32}{15}, &
\displaystyle\tilde\mathcal{D}=\pm\frac {\sqrt{65}\sqrt{561}}{11329956}
(26928P_0 + 8125),\\[2.7mm]
&\displaystyle\tilde\mathcal{B}=\pm\frac{8}{8415}\sqrt{65}\sqrt{561},
 &\displaystyle \tilde\mathcal{E}= -\:\frac{32}{5}
P_0^{2} - \frac{3496}{1683} P_0 - \frac{333125}{7553304},  \\[2.7mm]
&\displaystyle
\tilde\mathcal{C}= -\:\frac{32}{5}P_0 - \frac{1748}{1683}, \quad
&\displaystyle H=\frac{16}{15}P_0^{3} + \frac{7291}{13464}P_0^{2} +
\frac {6426875}{181279296} P_0 +
\frac{17551324375}{9762977765376}.\quad \\
\end{array}
\]

If the right-hand side of eq.~$(6)$ is a polynomial with multiple
  roots, then the function $y$ can be expressed in terms of elementary
  functions.  For example, at $P_0=0$ substitution of solutions 3-4
into eq.~$(5)$ gives
$$
  y=-\:\frac{5}{3\left(1-3\sin\left(\frac{t-t_0^{\vphantom{7}}}{3}
\right)\right)^{2^{\vphantom{27}}}},\eqno(12)
$$
where $t_0$ is an arbitrary constant.

From (11) we obtain the following values of $\mu$:
$$
\begin{array}{@{}ll@{}} 1.  &
 \mu=0,\\[2.7mm] 2. &
\mu=\frac{160}{1089}P_0^3 +\frac{680}{11979}P_0^2 -\frac
{800}{1185921}P_0-\frac{7000}{1056655611},\\[2.7mm]
3-4.&
\mu=\frac{4}{3}P_0^4 +\frac{5}{54}P_0^3+\frac{50}{729}P_0^2,\\[2.7mm]
5-6. &
 \mu={}-\frac{52}{561}P_0^4-\frac{81640}{944163}P_0^3-
\frac{4458460825}{152546527584}P_0^2 - 
\frac{539878421875}{128367902961936}P_0
-\frac{728473377734375}{6703885364284145664}.\\
\end{array}
$$

\subsection{Motion trajectories}

Let us consider the equations of the motion trajectories
at $C=-16/5$ and $\lambda=1/9$.
In the case of the solutions with $\tilde {\cal B}=\tilde {\cal D}=0$
the trajectory equation can be reduced either to $x^2=0$ (solution 1), or
to
$$
x^2 + \frac{6}{5}\left(y + \frac{20}{99}\right)^2=\frac{50}{1089}.\eqno(13)
$$
In the last case (solution 2) the motion trajectory is an ellipse.
Note, however, that the real motion does not necessarily affect the whole
ellipse: it depends on two arbitrary parameters. The energy $H$
can be considered as one of them.

In the case of solutions  3-4  the trajectory equation is the following:
$$
 \left(x^2 + \frac{5}{9}y\right)^2+\frac{5}{27}(y-P_0)(2y+P_0)^2 = 0.
\eqno(14)
$$
If $P_0=0$  (see $(12)$), the equation for one
of the trajectory branches entirely coincides with the equation
obtained in~\cite{Timosh}. The condition  $y<0$ is always required for
the existence of real motion along these trajectories. Formula $(9)$
describes precisely such a solution. For solutions 5-6 the trajectory
equation has the same form as for solutions  3-4.

\section{THREE--PARAMETER SOLUTIONS}

The Ablowitz--Ramani--Segur algorithm of the Painlev\'e
test~\cite{ARS} is very useful for obtaining the solutions as
formal Laurent series. Let the behavior of a solution in the
neighborhood of the singularity point $t_0$ be algebraic, i.e.,
$x$ and $y$ tend to infinity as some powers:
$x=a_{\alpha}(t-t_0)^\alpha$ and $y=b_{\beta}(t-t_0)^\beta$,
 where $\alpha$, $\beta$,  $a_{\alpha}$ and $b_{\beta}$   are some
constants.   If  $\alpha$  and  $\beta$ are negative integer
numbers, then substituting  the Laurent series expansions one
can transform nonlinear differential equations into a system of
linear algebraic equations on coefficients of Laurent series. If a
single-valued solution depends on more than one arbitrary
parameters then some coefficients of its Laurent series have to be
arbitrary and the corresponding systems have to have zero
determinants. The numbers of such systems (named {\it resonances}
or {\it Kovalevskaya exponents}) can be determined due to the Painlev\'e
test.

 Two possible dominant behaviors and resonance
structures of solutions of
the generalized H\'enon--Heiles system~\cite{Tabor,Melkonian} and eq.~$(3)$
are presented in the Table.

\vspace{7.2mm}
\noindent\begin{tabular}{|l|l|}
\hline  { }  { }
{ \it Case 1}  & { }  { }  {\it Case 2}: $\beta<\Re e(\alpha)$\\[3.7mm]
\hline { } $\displaystyle\alpha=-2$,&
 { }  $\displaystyle\alpha=\frac{1\pm\sqrt{1-48/C}^{\vphantom{7^4}}}{2}$,\\[2.7mm]
 { } $\displaystyle\beta=-2$, & { }  $\displaystyle\beta=-2$,\\[1mm]
 { } $\displaystyle a_{\alpha}=\pm 3\sqrt{2+C}$, &
 { } $\displaystyle a_{\alpha}=c_1^{\vphantom{4}}$ (arbitrary),\\[2.7mm]
 { } $\displaystyle b_{\beta}^{\vphantom{27}}=-3$,
&  { }  $\displaystyle b_{\beta}^{\vphantom{27}}=\frac{6}{C}$, \\[2.7mm]
 { } $\displaystyle r=-1,\; 6,\; \frac{5}{2}\pm\frac{\sqrt{1-24(1+C)}}{2}$
 & { }  $\displaystyle r=-1,\; 0,\; 6,\;
\mp\sqrt{1-\frac{48}{C}}$\\[3.7mm]
 { } $\displaystyle r_4^{\vphantom{7^4}}=-1,\; 10,\;
\frac{5}{2}\pm\frac{\sqrt{1-24(1+C)}}{2}$ { }   &
 { }  $\displaystyle r_4^{\vphantom{7^4}}=-1,\; 5,\;
5-\sqrt{1-\frac{48}{C}},\; 5+\sqrt{1-\frac{48}{C}}$ { }   \\[2.7mm] \hline
\end{tabular}
\vspace{7.2mm}

The values of $r$ denote resonances: $r=-1$ corresponds to
arbitrary parameter $t_0$; $r=0$ (in the {\it Case 2}) corresponds to
arbitrary parameter $c_1^{\vphantom{4}}$.
Other values of $r$ determine powers of $t$, to be exact,
$t^{\alpha+r}$ for $x$ and $t^{\beta+r}$ for $y$, at which new
arbitrary parameters can appear as solutions of the
linear systems with zero determinant.
Note, that the dominant behaviour and the resonance structure depend
only on $C$.

It is necessary for the integrability of system (2) that all values of
$r$ be integer and that all systems with zero determinants
have solutions for any values of the free parameters entering these
systems. This is possible only in the integrable cases (i)--(iii).

 For the search for special solutions, it is interesting to
consider such values of $C$, for which $r$ are integer
numbers either only in {\it Case 1} or only in {\it Case 2}.
If there exist a negative integer resonance, different from $r=-1$,
then such Laurent series expansion corresponds rather to special
than general solution~\cite{Tabor}.
We demand that all values of $r$, but one, are nonnegative integer numbers
and all these values are different. From these conditions we obtain the
 following values of $C$: \ $C=-1$ and $C=-4/3$ ({\it Case 1}), or
$C=-16/5$, $C=-6$ and $C=-16$ ({\it Case 2},
$\alpha=\frac{1-\sqrt{1-48/C}^{\vphantom{7^4}}}{2}$), and also
$C=-2$, in which these two {\it Cases} coincide. It is remarkable that
only for these values of $C$ there exist solutions of system  $(8)$
with ${\tilde\mathcal{B}}\neq 0$ or ${\tilde\mathcal{D}}\neq 0$.

Let us consider the possibility of existence of the single-valued
three-parameter solutions in all these cases.  To obtain the result
for an arbitrary value of $\mu$, we consider eq.~$(3)$ with
an arbitrary $H$.  Note, that the values  of resonances obtained from
eq.~$(3)$ (in the Table they are signified as $r_4$)
are different from $r$, but we obtain the same result:
condition that all values of $r_4$, but $r_4=-1$, are nonnegative integer
numbers gives the same values of $C$.

At $C=-2$ we have a contradiction: $r_4=0$, but $b_{-2}$ is not
arbitrary parameter: $b_{-2}=-3$.
This is the consequence of the fact
that, contrary to our assumption, the behaviour of the general solution in the
neighborhood of a singular point is not algebraic, because its dominant
term includes logarithm~\cite{Tabor}.  At $C=-6$ and any value of
other parameters the exact four-parameter solutions are known.
In cases $C=-1$ and $C=-16$ the substitution 
of an unknown function as the Laurent
series leads to the conditions $\lambda_1 =
\lambda_2$ or $\lambda_1 = \lambda_2/16$ accordingly. 
Hence, in nonintegrable cases
three-parameter local solutions have to include logarithmic
terms.  Single-valued three-parameter solutions can exist only in two
above-mentioned nonintegrable cases: $C = -16/5$ and $C = -4/3$.  

Using the method of construction of the Laurent series solutions
for nonlinear differential equations describing in~\cite{VernovTMF},
we obtain single-valued local solutions of eq.~$(3)$ both at
$C=-16/5$ and at $C=-4/3$. Values of other parameters are arbitrary.

At $C=-4/3$ these solutions are:
$$
\begin{array}{@{}l@{}}
\displaystyle  y=-3\frac{1}{t^2}+b_{-1}\frac{1}{t}+
 \frac{29}{24}b_{-1}^{2}+\frac{1}{2}\lambda_1-\frac{3}{4}\lambda_2+
 \left(\frac{17}{6} b_{-1}^{2}+\frac{5}{3}\lambda_1-\frac{5}{4}
 \lambda_2 \right) b_{-1}t+b_2t^2- {}\\[4.7mm]
\displaystyle {} - \left(\frac {55}{12} \lambda_1b_{-1}^2
+\frac { 131}{90}\lambda_1^2 +\frac {33}{40} \lambda_2^2 +
\frac{9359}{2592}b_{-1}^4+b_2 -\frac{55}{16}
\lambda_2b_{-1}^2 -\frac{131}{60}\lambda_1\lambda_2\right)
b_{-1}^2t^3+\dots.\\[2.7mm]
\end{array}  \eqno(15)
$$
There exist four possible values of the parameter $b_{-1}$:
$$
b_{-1}=\pm\sqrt{\frac{ 105\lambda_2-
140\lambda_1 +\sqrt{7(1216\lambda_1^2-1824\lambda_1\lambda_2
+783\lambda_2^2)} }{385}}
$$
or
$$
b_{-1}=\pm\sqrt{\frac{105\lambda_2 - 140\lambda_1 
-\sqrt{7(1216\lambda_1^2-1824\lambda_1
\lambda_2+783\lambda_2^2)}}{385}}.
$$
The parameters $b_2$ and $b_8$, coefficients at $t^2$ and
$t^8$ correspondingly, are arbitrary.
The energy $H$ enters in coefficients beginning from $b_4$.

At $C=-16/5$ we obtain the following solutions:
$$
\begin{array}{@{}l@{}} \displaystyle
y= {} - \frac{15}{8t^{-2}}+\tilde b_{-1}- \frac{5}{32}\lambda_2
+\frac{62}{45} \tilde b_{-1}^2+
 \left(\frac{5}{12}\lambda_1 + \frac{632}{225}
\tilde b_{-1}^2-\frac {25}{192}\lambda_2  \right)\tilde  b_{-1} t+
{}\\[4.7mm]
\displaystyle
{} + \left(\frac{29}{15}\lambda_1\tilde b_{-1}^2-\frac{1}{
128}\lambda_2^{2} -\frac{29}{48}\lambda_2\tilde b_{-1}^2
+\frac{102272}{10125}\tilde b_{-1}^4 \right) t^2
+\tilde b_3t^3 +\dots, \end{array} \eqno(16)
$$          
with
$$
\tilde
b_{-1}=\frac{{}\pm 3}{41888}
\sqrt{6872250\lambda_2-21991200\lambda_1+
52360\sqrt{71680\lambda_1^2-44800\lambda_1\lambda_2
+13545\lambda_2^2}}
$$
or
$$
\tilde
b_{-1}={} \frac{{}\pm 3}{41888}
\sqrt{6872250\lambda_2-21991200\lambda_1 -
52360\sqrt{71680\lambda_1^2-44800\lambda_1\lambda_2
+13545\lambda_2^2}}.
$$

The coefficients  $\tilde b_3$ and $\tilde b_8$  are arbitrary
parameters. Beginning from $\tilde b_4$ some coefficients
include the energy $H$. So, the obtained local solutions depend on
four independent parameters:
$t_0$, $H$ and two coefficients ($b_2$ and $b_8$ or
$\tilde b_3$ and $\tilde b_8$).

We have found local single-valued solutions.
Of course, existence of local single-valued solutions is
necessary, but not sufficient condition to exist global ones,
because solutions, which are single-valued in the  neighborhood of
one singularity point, can  be multivalued in the  neighborhood of
another singularity point. So, we can only assume that global three-parameter
solutions are single-valued. If we assume this and moreover that
these solutions are elliptic functions (or some degenerations of
them), then we can seek them as solutions of some polynomial first
order equations. There are a few methods to construct such
solutions~\cite{Weiss1,CoMu92,Cantos,Fan}. Using these methods one
represents a solution of a nonlinear ordinary differential
equation (ODE) as the finite Taylor or Laurent series of
elliptic functions or degenerate elliptic functions, for example,
$\tanh(t)$. Similar method is applied in this paper to find
two-parameter solutions. These methods use results of the
Painlev\'e test, but don't use the obtained Laurent-series
solutions. In 2003 R. Conte and M. Musette~\cite{CoMu03} have
proposed the method, which uses such solutions.

The classical theorem, which was established by
Briot and Bouquet~\cite{BriBo}, proves that if the general solution of a
polynomial autonomous first order ODE is single-valued, then this
solution is either an elliptic function,  or a rational function of
$e^{\gamma x}$, $\gamma$ being some constant, or a rational function of
$x$. Note that the third case is a degeneracy of the second one, which in
its turn is a degeneracy of the first one.  It has been
proved by Painlev\'e~\cite{Painleve} that the necessary form of the
polynomial autonomous first order ODE with the
single-valued general solution is
$$
\sum_{k=0}^{m}
\sum_{j=0}^{2m-2k}h_{jk}^{\vphantom{27}}\: y^j y_{t}^k=0,
\qquad h_{0m}^{\vphantom{27}}=1, \eqno(17)
$$
in which $m$ is a positive integer number and $h_{jk}$ are constants.

Rather than to substitute eq.~$(17)$ in some nonintegrable system,
one can substitute the Laurent series of unknown special solutions, for
example, $(15)$ or $(16)$
in eq.~$(17)$ and obtain a system, which is linear in
$h_{jk}^{\vphantom{27}}$ and nonlinear in the parameters, including in the 
Laurent coefficients~\cite{CoMu03}. There are a few computer algebra 
algorithms which allow to obtain this system from the  given  Laurent 
series.  Moreover it is possible to exclude  all $h_{jk}^{\vphantom{27}}$ 
from this system and obtain a nonlinear system in parameters of 
nonintegrable system and free parameters from the Laurent series.  The 
main preference of this method is that the number of unknowns in the 
resulting nonlinear algebraic system does not depend on number of 
coefficients of the first order equation.  For example,  eq. $(17)$ with 
$m=8$ includes 60 unknowns $h_{jk}^{\vphantom{27}}$, and it is not 
possible use the traditional way to find similar solutions. Using this 
method we always obtain nonlinear system in 5 variables: $\lambda_1$, 
$\lambda_2$, $H$ and two arbitrary coefficients of the Laurent-series 
solutions. We hope that this method allows us to find three-parameter 
global solutions.

\section{Conclusions}
Two nonintegrable cases ($C=-16/5$ or $C=-4/3$, 
$\lambda_1$, $\lambda_2$ and $\mu$ are arbitrary) of the generalized 
H\'enon--Heiles system with the nonpolynomial term have been considered.  
To avoid problems with the nonpolynomial term we have transformed system 
into the fourth-order equation.  Two-parameter elliptic solutions for this 
equation have been found in both above-mentioned cases.  Two different 
solutions correspond to each pair of parameter values. The Painlev\'e 
test does not show any obstacle to the existence of three-parameter 
single-valued solutions, so,  the probability to find exact, for example 
elliptic, three-parameter solutions, that generalize the obtained 
solutions, is high.

S.Yu.V. is grateful to  \  F.~Calogero, \  R.~Conte,
 \  V.~F.~Edneral \ and  \  A.~K.~Pogrebkov \
for valuable discussions. This work has been supported by
Russian Fede\-ration President's Grants NSh--1685.2003.2
and NSh--1450.2003.2
and by the grant of  the scientific Program
"Universities of Russia".

\section*{APPENDIX}

In two nonintegrable cases 
($C=-16/5$ and $C=-4/3$) for arbitrary $\lambda_1$ and $\lambda_2$ we 
obtain that six solutions of system~$(8)$ correspond to each value of 
$P_0$.  Two of them (with $\tilde\mathcal{B}=\tilde\mathcal{D}=0$) 
generate solutions of eq.~$(4)$.
Other solutions of system~$(8)$ can be separated on
pairs such as each pair of solutions corresponds to one
two-parameter function $y=\varrho^2+P_0$, where $\varrho$
satisfies eq.~$(6)$ with the following values of coefficients:

\[
\noindent
\begin{array}{@{}l}
\displaystyle C=-\:\frac{16}{5},   \\[4.7mm]
\displaystyle \tilde\mathcal{A}={-\:\frac{32}{15}}, \\[4.7mm]
\displaystyle
\tilde\mathcal{B}=-\: \frac{\sqrt{1122}(1120 \lambda_1  + 41888 P_0
 + 65 S_q + 6195\lambda_2)\sqrt{F_1(\lambda_1,\lambda_2, P_0)}}
{29373960(3600 \lambda_1 ^{2} - 1120
\lambda_1 P_0 - 2425 \lambda_1\lambda_2 - 20944 P_0^{2} - 6195\lambda_2 P_0 +
 225\lambda_2^2)}, \\[4.7mm]
 \displaystyle \tilde\mathcal{C}= -
\frac{240}{187}\lambda_1 - \frac{32}{5} P_0 + \frac{4}{1309}
S_q - \frac{112}{187}\lambda_2, \\[4.7mm]
 \displaystyle \tilde\mathcal{D}=
\frac{\sqrt{1122}}{5874792}\sqrt{F_1(\lambda_1,\lambda_2,P_0)},\\[4.7mm]
\displaystyle \tilde\mathcal{E}= 
\frac{88320}{244783}\lambda_1^{2}
-  \frac {480}{187}\lambda_1 P_0 + \frac {885}{244783}\lambda_1
S_q - \frac{153375}{244783} \lambda_1\lambda_2  - \mbox{} 
  \\[3.7mm]
\displaystyle  \mbox{} -
\frac {32}{5}P_0^{2} + \frac {8}{1309}P_0S_q
  - \frac{224}{187}\lambda_2P_0 - 
\frac{685}{3916528}\lambda_2S_q + 
\frac{168855}{3916528}\lambda_2^2,   \\[4.7mm]
\displaystyle H= -
 \frac {11516270}{45774421}  \lambda_1^{3} +
 \frac {8740}{34969} \lambda_1^{2} P_0
  -  \frac{3296515}{2563367576}  \lambda_1^{2}
S_q +  \frac{50336425}{183097684}\lambda_1 ^{2}\lambda_2
+ \mbox{} \\[3.7mm]
\displaystyle  \mbox{} + \frac {258}{187}\lambda_1 P_0^{2}  -
\frac{8209}{1958264}\lambda_1
P_0S_q  +  \frac{76915}{279752}  \lambda_1\lambda_2 P_0
+  \frac{12202395}{82027762432}\lambda_1\lambda_2 S_q 
-  \mbox{} \\[3.7mm]
\displaystyle  \mbox{}  -
\frac{131879855}{11718251776}\lambda_1\lambda_2^2
 - \frac{43}{13090}P_0^{2}S_q +
 \frac {103}{1496}\lambda_2 P_0^{2} +
  \frac {8881}{31332224}
\lambda_2  P_0S_q -  \mbox{}\\[3.7mm]
\displaystyle  \mbox{}  - \frac {71205}{4476032}
\lambda_2^2 P_0 - \frac {12990165}{1312444198912}\lambda_2^2 S_q -
 \frac {168661575}{187492028416}\lambda_2^3   
+ \frac {16}{15}P_0^{3},\\[7.2mm]
\end{array}
\]
\[
\begin{array}{@{}l}
\displaystyle C=-\:\frac{4}{3}, \qquad
  \tilde\mathcal{A}=-\:\frac{4}{3}, \\[3.7mm]
\displaystyle\tilde\mathcal{B}=\frac{\sqrt{330}(952
\lambda_1  - 616 P_0 + 13 R_q 
- 945\lambda_2)\sqrt{F_2(\lambda_1,\lambda_2, P_0)}}{38115(
432 \lambda_1 ^{2} + 952 \lambda_1  P_0 - 291 \lambda_1\lambda_2  -
308 P_0^{2} - 945 P_0\lambda_2 + 27\lambda_2^2)}, 
\\[3.7mm] \displaystyle\tilde\mathcal{C}=- \frac{4}{33}\lambda_1  - 4 P_0 
-  \frac {1}{66} R_q - \frac {31}{22}\lambda_2,\\[3.7mm] 
\displaystyle\tilde\mathcal{D}=\frac{\sqrt{330}}{7623}
\sqrt{F_2(\lambda_1,\lambda_2,P_0)},\\[3.7mm]
\displaystyle\tilde\mathcal{E}=\frac{3394}{363}\lambda_1^2 +
\frac{54}{11}\lambda_1 P_0 -
 \frac {1123}{10164}  \lambda_1  R_q -
 \frac {5897}{484}  \lambda_1\lambda_2 - \mbox{} \\[2.7mm]
\displaystyle
\mbox{} - { \frac {17}{3}}  P_0^{2} -
 \frac {31}{308}  P_0 R_q -
 \frac {349}{44}\lambda_2P_0 + \frac {1223}{27104}\lambda_2  R_q +
 \frac {13005}{3872}\lambda_2^2,\\[3.7mm]
 \displaystyle H= - { \frac {552922}{83853}}
 \lambda_1 ^{3} - { \frac {29801}{2541}}  \lambda_1 ^{2}
P_0 + { \frac {173605}{2347884}}  \lambda_1
^{2} R_q + {\frac{778033}{74536}}
\lambda_1^{2}\lambda_2\lambda_2 - { \frac {185}{66}}  \lambda_1
P_0^{2} +\mbox{} \\[2.7mm]
\displaystyle
\mbox{} + \frac{3001}{20328}  \lambda_1
 P_0 R_q+ \frac{104959}{6776} \lambda_1\lambda_2
P_0 - { \frac {695609}{12522048}}\lambda_1 \lambda_2
 R_q -  \frac{2990049}{596288} \lambda_1\lambda_2^2
  +\frac {89}{1232}  P_0^{2} R_q
  +  \mbox{}
\\[2.7mm]
\displaystyle\mbox{} + \frac {5}{2}  P_0^{3}+  \frac {865}{176} \lambda_2
 P_0^{2} - \frac{3065}{54208}\lambda_2
 P_0
R_q - \frac {225909}{54208}\lambda_2^2 P_0
 +  \frac {2733}{260876} \lambda_2^2 R_q +
 \frac {57699}{74536}\lambda_2^3,\\
\end{array}
\]
where
$$
\begin{array}{@{}l}
\displaystyle F_1(\lambda_1,\lambda_2,P_0)\equiv 
39474176000\lambda_1^{3} +  
 122782105600 \lambda_1^{2} P_0 
- 104358400 \lambda_1^{2} S_q - \mbox{} \\[2.7mm]   
\displaystyle \mbox{}  -
17822336000 \lambda_1^{2}\lambda_2 + 210552545280 \lambda_1 P_0^{2} 
- 680261120 \lambda_1  P_0 S_q
- 10941145600 \lambda_1\lambda_2  P_0 
-\mbox{}
 \\[2.7mm]  \displaystyle \mbox{} -  41066800 \lambda_1 \lambda_2
 S_q  + 8305290000 \lambda_1\lambda_2^2 - 
501315584 P_0^{ 2} S_q  
- 65797670400\lambda_2 P_0^{2} +\mbox{} \\[2.7mm]  \displaystyle 
\mbox{}+ 
55920480 P_0 S_q+ 1611640800\lambda_2^2 P_0 + 2884725\lambda_2^2 
S_q - 468507375\lambda_2^3,\\[2.7mm] \displaystyle 
S_q\equiv\pm\sqrt{35(2048\lambda_1^2 - 
1280\lambda_1\lambda_2+ 387\lambda_2^2)},\\[2.7mm] \displaystyle 
F_2(\lambda_1,\lambda_2,P_0)\equiv 
2099776\lambda_1^3 - 497728\lambda_1^2 P_0 - 20008 \lambda_1 ^{2} 
R_q - 
4911144 \lambda_1^2\lambda_2 + 948640 \lambda_1  P_0^{2} +\mbox{}\\[2.7mm] 
\displaystyle
\mbox{}+ 19096 \lambda_1  P_0
 R_q + 1458072 \lambda_1\lambda_2  P_0  
+ 37173 \lambda_1\lambda_2  
R_q + 3943233\lambda_1\lambda_2^2  + 6776 P_0^{2} 
 R_q-\mbox{}\\[2.7mm] \displaystyle \mbox{} - 711480 
 \lambda_2P_0^{2} - 9240\lambda_2 P_0  R_q
 - 615384\lambda_2^2 P_0 - 13581 
\lambda_2^2R_q - 1006425\lambda_2^3, \\[2.7mm] 
\displaystyle
 R_q\equiv\pm\sqrt{7(1216\lambda_1^2 - 
1824\lambda_1\lambda_2 + 783\lambda_2^2)}.  
\end{array} 
$$

\end{document}